\documentclass{pnastwo}

\usepackage{graphicx}
\usepackage{times}

\usepackage{amssymb,amsfonts,amsmath}

\newcommand{\apj}{    {\it Astrophysical Journal}}
\newcommand{\aap}{    {\it Astronomy and Astrophysics}}
\newcommand{\solphys}{    {\it Solar Physics}}
\newcommand{\mnras}{    {\it Monthly Notices of the Royal Astronomical Society}}
\newcommand{\apjl}{    {\it Astrophysical Journal Letters}}
\newcommand{\nat}{    {\it Nature}}

\contributor{Submitted to Proceedings
of the National Academy of Sciences of the United States of America}
\url{www.pnas.org/cgi/doi/10.1073/pnas.0709640104}
\copyrightyear{2008}
\issuedate{Issue Date}
\volume{Volume}
\issuenumber{Issue Number}
\begin{document}
% Include your paper's title here

\title{ANOMALOUSLY WEAK SOLAR CONVECTION}% AND ANOMALOUSLY WEAK TRANSPORT VELOCITIES IN THE CONVECTION ZONE}
%% make sure you have the nature.cls and naturemag.bst files where
%% LaTeX can find them
%\usepackage{graphicx}
%\bibliographystyle{naturemag}
%\usepackage{amssymb}

%% Notice placement of commas and superscripts and use of &
%% in the author list

\author{Shravan M. Hanasoge\affil{1}{Department of Geosciences, Princeton University, NJ 08544, USA}\affil{2}{Max-Planck-Institut f\"{u}r Sonnensystemforschung, 37191 Katlenburg-Lindau, Germany} \and
Thomas L. Duvall, Jr.\affil{3}{Solar Physics Laboratory, NASA/Goddard Space Flight Center, MD 20771, USA}
\and
 Katepalli~R. Sreenivasan\affil{4}{Courant Institute of Mathematical Sciences, New York University, NY 10012, USA}
%\normalsize{$^\ast$To whom correspondence should be addressed; E-mail:  hanasoge@princeton.edu.}
}

\contributor{Submitted to Proceedings of the National Academy of Sciences
of the United States of America}

%%%%%%%%%%%%%%%%% END OF PREAMBLE %%%%%%%%%%%%%%%%

\maketitle

\begin{article}

\begin{abstract}
Convection in the solar interior is thought to comprise structures on a spectrum of scales. This
conclusion emerges from phenomenological studies and numerical simulations, though neither covers
the proper range of dynamical parameters of solar convection.
Here, we analyze observations of the wavefield in the solar photosphere using
techniques of time-distance helioseismology to image flows in the solar interior.
We downsample and synthesize 900 billion wavefield observations to produce 3 billion cross-correlations, which we average and fit, measuring 5 million wave travel times.
Using these travel times, we deduce the underlying flow systems and study their statistics to bound convective
velocity magnitudes in the solar interior, as a function of depth and
spherical-harmonic degree $\ell$. Within the wavenumber band $\ell<60$, Convective velocities are 20-100 times weaker than current theoretical estimates.
This suggests the prevalence of a different paradigm of turbulence from that predicted by existing models, prompting the question:
what mechanism transports the heat flux of a solar luminosity outwards?
Advection is dominated by Coriolis forces for wavenumbers $\ell<60$, with Rossby numbers smaller than $\sim10^{-2}$ at $r/R_\odot=0.96$, 
suggesting that the Sun may be a much faster rotator than previously thought, and that large-scale convection may be quasi-geostrophic.
The fact that iso-rotation contours in the Sun are not co-aligned with the axis of rotation suggests the presence of a latitudinal entropy gradient.
\end{abstract}
%The ``rapidity" of interior solar rotation, characterized via
%the Rossby number as the ratio of advection to Coriolis forces, is presently constrained only through simulations and thought to be comparable to unity
%for all but the largest scales. 

\section{Introduction}
The thin photosphere of the Sun, where thermal transport is dominated by free-streaming radiation, shows a spectrum in which granulation and supergranulation are most prominent. Observed properties of granules, such as spatial scales, radiative intensity and photospheric spectral-line formation are successfully reproduced by numerical simulations \cite{stein00,voegler2005}. In contrast, convection in the interior is not directly observable and likely governed by aspects more difficult to model, such as the integrity of descending plumes to diffusion and various instabilities \cite{rast98}. Further, solar convection is governed by extreme parameters \cite{miesch05} (Prandtl number $\sim 10^{-6} - 10^{-4}$, Rayleigh number $\sim 10^{19} - 10^{24}$, and Reynolds number $\sim 10^{12} - 10^{16}$), which make 
fully resolved three-dimensional direct numerical simulations impossible for the foreseeable future. It is likewise difficult to reproduce them in laboratory experiments.

Turning to phenomenology, mixing-length theory (MLT) is predicated on the assumption that parcels of fluid of a specified spatial and velocity scale transport heat over one length scale (termed the {\it mixing length}) and are then mixed in the new environment. While this picture is simplistic \cite{cox_2004}, it has been successful in predicting aspects of solar structure as well as the dominant scale and magnitude of observed surface velocities. MLT posits a spatial convective scale that increases with depth (while velocities reduce) and coherent large scales of convection, termed {\it giant cells}. Simulations of anelastic global convection \cite{miesch_etal_08,charbonneau10,kapyla1, kapyla2}, more sophisticated than MLT, support the classical picture of a turbulent cascade. 
The ASH simulations \cite{miesch_etal_08} solve the non-linear compressible Navier-Stokes equations in the anelastic limit, i.e., where acoustic waves, which oscillate at very different timescales, are filtered out.
Considerable effort has been spent in attempting surface \cite{hathaway00} and interior detection \cite{duvall, duvall03} of giant cells, but evidence supporting their existence has remained inconclusive.

\section{Results}
\begin{figure}
\centering{\includegraphics[width=\linewidth]{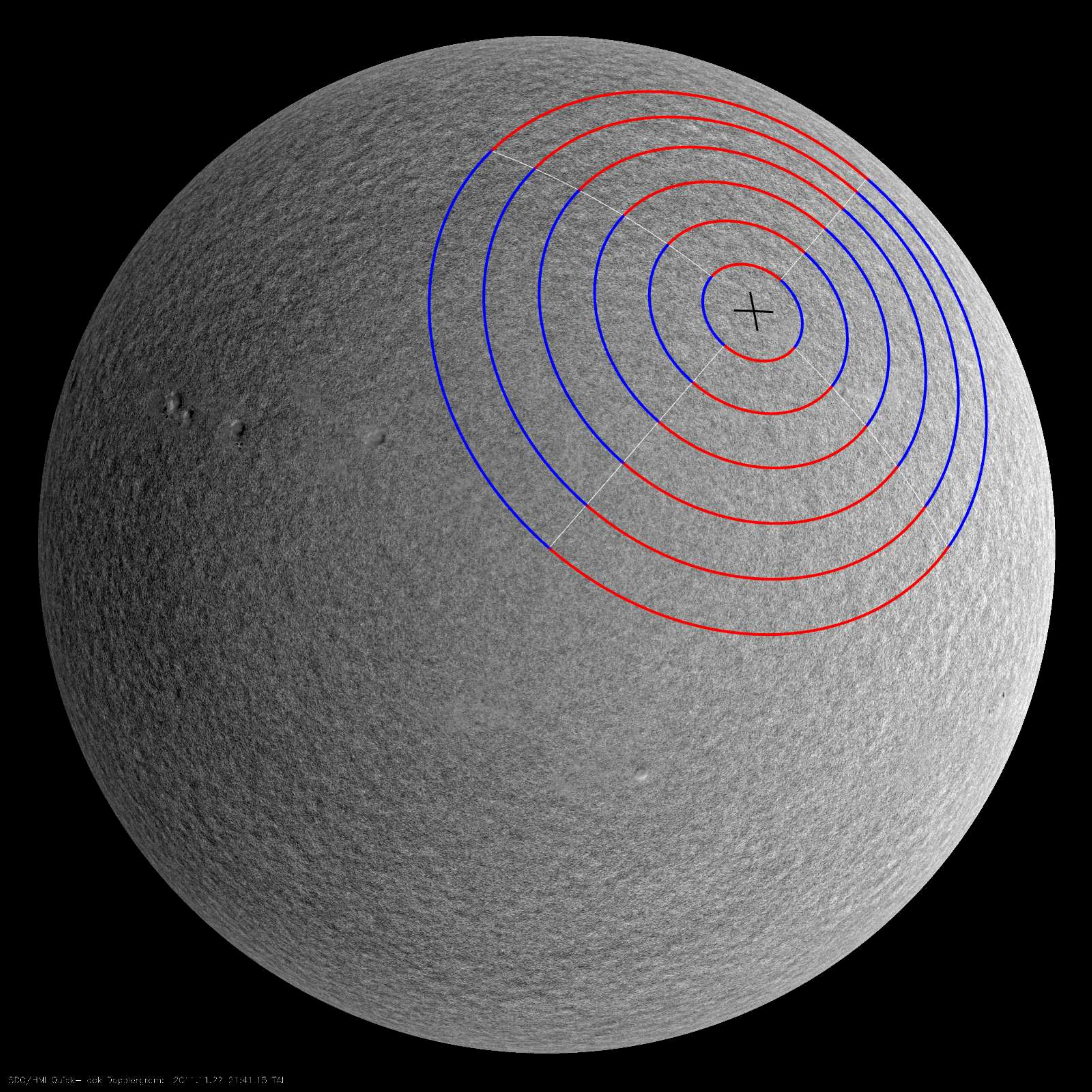}}\label{hmifig}
\caption{
Line-of-sight Doppler velocities are measured every 45 seconds at $4096\times4096$ pixels on the solar photosphere by the Helioseismic and Magnetic Imager (background image).
We cross correlate wavefield records of temporal length $T$ at points on opposing quadrants (blue with blue or red with red). These ``blue" and ``red" correlations are separately averaged,
respectively sensitive to longitudinal and latitudinal flow at $(\theta, \phi; r/R_\odot = 0.96)$, where $(\theta,\phi)$ is the central point marked by a cross (see Figure~\ref{rays} for further illustration).
%The difference between travel times of waves propagating from one quadrant to the opposing one and is sensitive to the flow .
The longitudinal measurement is sensitive to flows in that direction while the latitudinal measurement to flows along latitude.
We create a travel-time maps $\delta\tau(\theta,\phi,T)$ by making this measurement about
various central points $(\theta,\phi)$ on the surface. Each travel time is obtained upon correlating the wavefield between 600 pairs of points distributed in azimuth.
}
\end{figure}

\begin{figure}
\centering{\includegraphics[width=\linewidth]{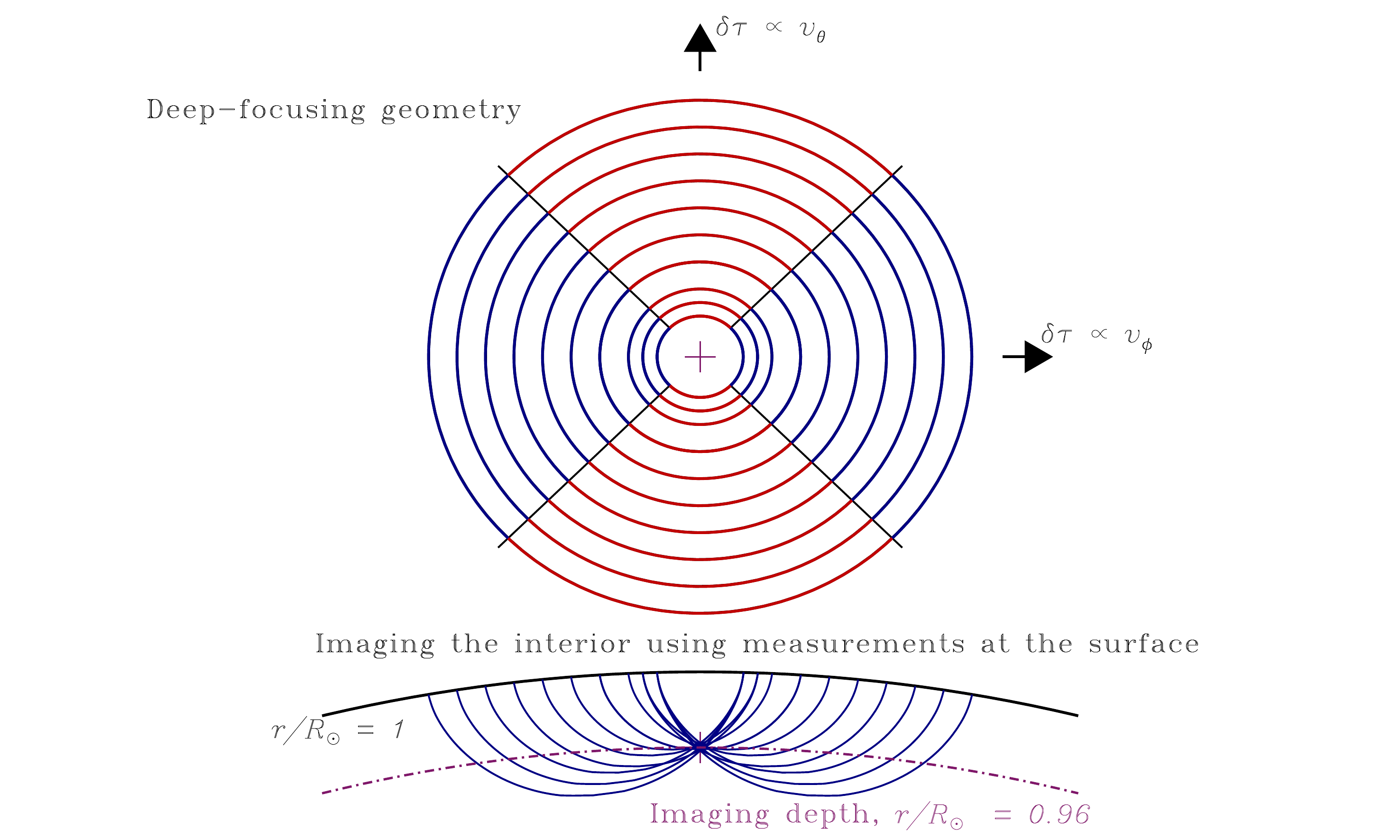}}\label{rays}
\caption{
The cross-correlation measurement geometry (upper panel; arrowheads - horizontal: longitude, and vertical: latitude) used to image the layer $r/R_\odot = 0.96$ (dot-dashed line). Doppler velocities of temporal length $T$ measured at the solar surface are cross correlated between point pairs at opposite ends of annular discs (colored red and blue); e.g., points on the innermost blue sector on the left are correlated with diagonally opposite points on the outermost blue sector on the right. Six-hundred correlations are prepared and averaged for each travel-time measurement.
Travel times of waves that propagate along paths in the direction of the horizontal and vertical arrows are primarily sensitive to longitudinal and latitudinal flows, $v_\phi$ and $v_\theta$, respectively. The focus point of these waves is at $r/R_\odot = 0.96$ (lower panel) and the measured travel-time shift $\delta\tau(\theta,\phi,T)$ is linearly related to the flow component $v(r/R_\odot = 0.96, \theta,\phi)$ with a contribution from the incoherent wave noise. We are thus able to {\it map} the flow field at specific depths $v(r,\theta,\phi)$ through appropriate measurements of $\delta\tau(\theta,\phi,T)$. For the inversions here, we create travel-time maps of  size $128 \times 128$ (see Figure~\ref{tt.map}). For reference, we note that the base of the convection zone is located at $r/R_\odot = 0.71$ and
the near-surface shear layer extends from $r/R_\odot = 0.9$ upwards.}
\end{figure}

\begin{figure}%\vspace{-3cm}
\centerline{\includegraphics[width=\linewidth]{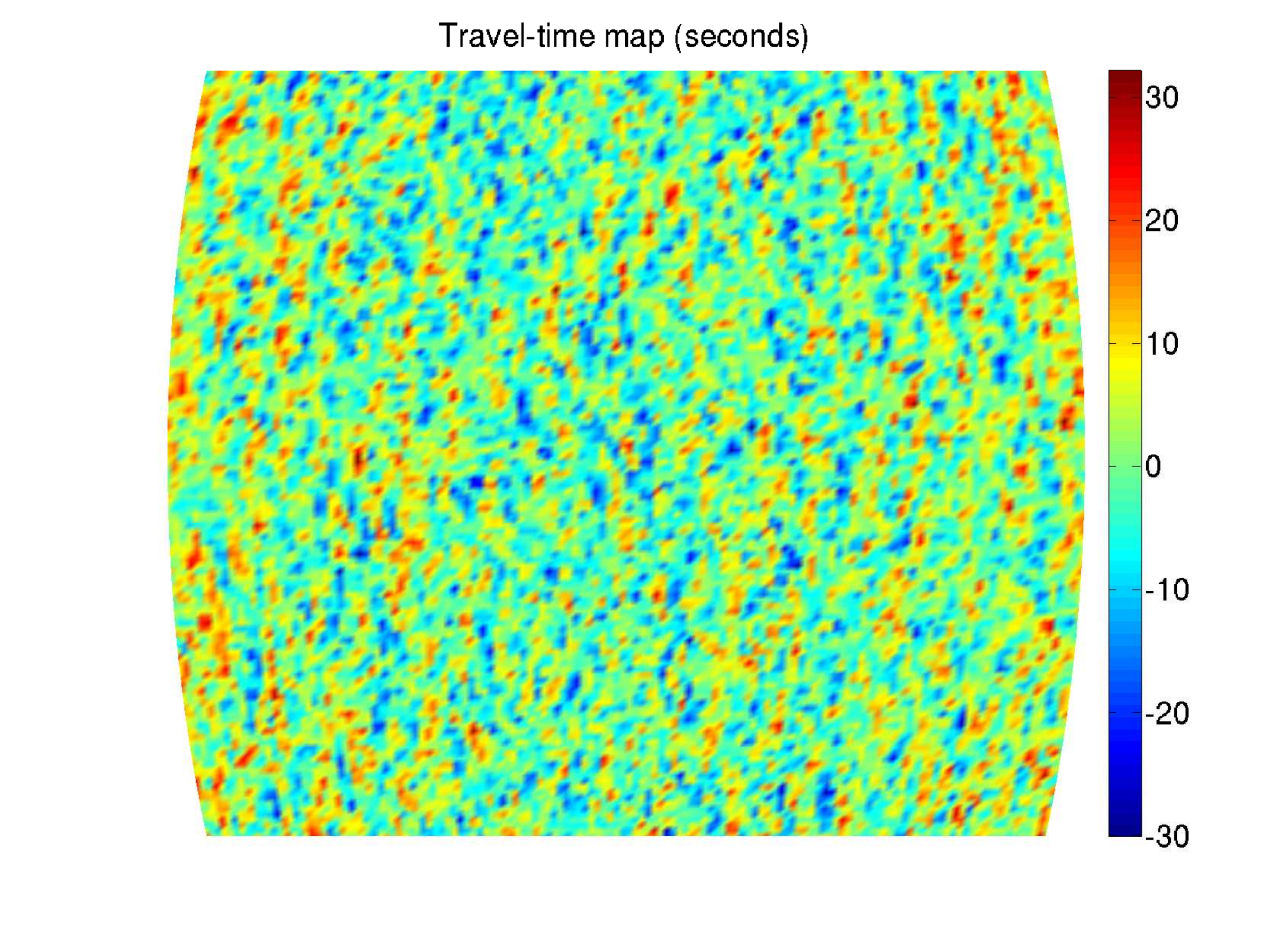}}%\vspace{-2.5cm}
\caption{
A travel-time map consisting 16,384 travel-time measurements, spanning a $60^\circ \times 60^\circ$ region (at a resolution of 0.46875 degrees per pixel) around the solar disk center, obtained by analyzing one day's worth of data taken by the Helioseismic and Magnetic Imager instrument \cite{hmi} onboard the Solar Dynamics Observatory satellite. 3.2 billion wavefield measurements were analyzed to generate 10 million correlations, which were averaged and fitted to generate this travel time map.
This geometry and these particular wave times are so chosen as to be sensitive to flow systems in the solar interior. The spectrum of these travel times shows no interesting or anomalous peaks that meet the detection criteria (described subsequently). 
}
\label{tt.map}
\end{figure}

Here, we image the solar interior using time-distance helioseismology \cite{duvall, duvall03, gizon2010}.
Raw data in this analysis are line-of-sight photospheric Doppler velocities measured by the Helioseismic and Magnetic Imager  \cite{hmi} onboard the Solar Dynamics Observatory. Two-point correlations from temporal segments of length $T$ of the observed Doppler wavefield velocities are formed and spatially averaged according to a deep-focusing geometry \cite{hanasoge10} (Figures~\ref{hmifig} and~\ref{rays}). We base the choice of $T$ on estimates of convective coherence timescales \cite{spruit74, gough77, miesch_etal_08}. These correlations are then fitted to a reference Gabor wavelet function \cite{duvall97} to obtain travel-time shifts $\delta\tau(\theta,\phi, T)$, where $(\theta, \phi)$ are co-latitude and longitude on the observed solar disk. By construction, these time shifts are sensitive to different components of 3D vector flows, i.e., longitudinal, latitudinal or radial, at specific depths of the solar interior ($r/R_\odot = 0.92, 0.96$) and consequently, we denote individual flow components (longitudinal or latitudinal) by scalars. Each point $(\theta,\phi)$ on the
travel-time map is constructed by correlating 600 pairs of points on opposing quadrants. A sample travel-time map is shown in Figure~\ref{tt.map}.

Waves are stochastically excited in the Sun, because of which the above correlation and travel-time measurements include components of incoherent wave noise, whose variance \cite{gizon_04} diminishes as $T^{-1}$. The variance of time shifts induced by convective structures that retain their coherence over timescale $T$ does {\it not} diminish as $T^{-1}$, allowing us to distinguish them from noise. We may therefore describe the total travel-time variance $\sigma^2(T) \equiv \sum_{\theta, \phi} \langle\delta\tau^2(\theta, \phi,T)\rangle $ as the sum of variances of signal $S^2$ and noise $N^2/T$, assuming that $S$ and $N$ are statistically independent. Angled brackets denote ensemble averaging over measurements of $\delta\tau(\theta,\phi,T)$ from many independent segments of temporal length $T$. Given a coherence time $T_{\rm coh}$, we fit $\sigma^2(T) = S^2 + N^2/T$ over $T<T_{\rm coh}$ to obtain the integral upper limit $S$. The fraction of the observed travel-time variance that cannot be modeled as uncorrelated noise is therefore $S^2/\sigma^2(T_{\rm coh})$. For averaging lengths $T_{\rm coh}$ (= 24 and 96 hours) considered here, we find this signal to be small, i.e., $S^2 \ll N^2/T_{\rm coh}$, which leads us to conclude that large-scale convective flows are weak in magnitude. Further, since surface supergranulation contributes to $S$, our estimates form an upper bound on ordered convective motions.

\begin{figure}
\centering{\includegraphics[width=\linewidth]{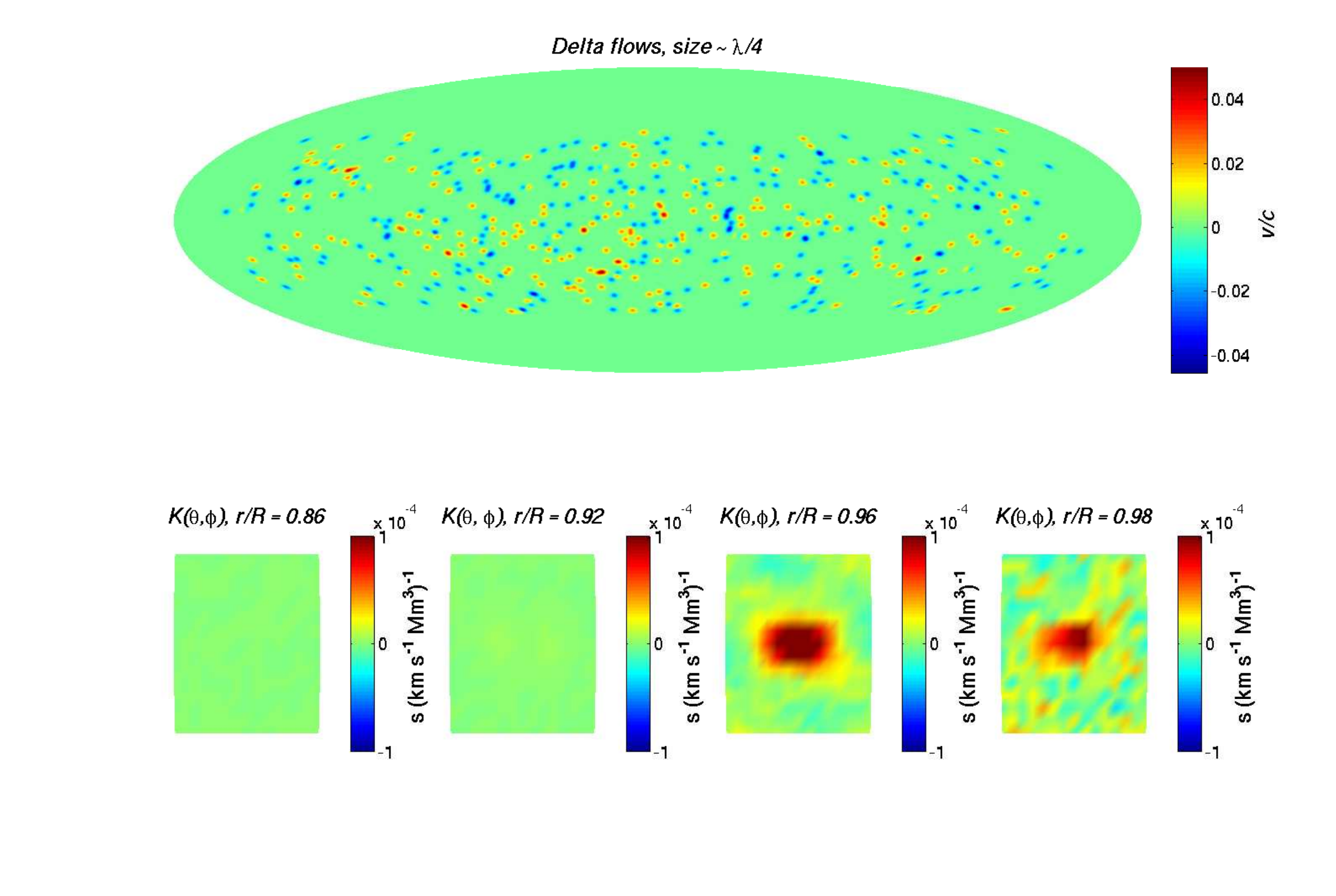}}\label{kernels}
\caption{
Because wavelengths of helioseismic waves may be comparable to or larger than convective features through which they propagate, the ray approximation is inaccurate and finite-wavelength effects must be accounted for when modeling wave propagation in the Sun \cite{dahlen99}. In order to derive the 3D finite-frequency sensitivity function (kernel) associated with a travel-time measurement \cite{duvall06}, we simulate waves propagating through a randomly scattered set of 500 east-west-flow `delta' functions, each of which is assigned a random sign so as not to induce a net flow signal \cite{hanasoge07} (upper panel). We place these flow deltas in a latitudinal band of extent 120$^\circ$ centered about the equator, because the quality of observational data degrades outside of this region. We perform six simulations, with these deltas placed at a different depth in each instance, so as to sample the kernel at these radii. The bottom four panels show slices at various radii of the sensitivity function for the measurement which attempts to resolve flows at $r/R_\odot = 0.96$. Measurement sensitivity is seen to peak at the focus depth, a desirable quality, but contains near-surface lobes as well. Note that the volume integral of flows in the solar interior with this kernel function gives rise to the associated travel-time shift, which explains the units.}
\end{figure}

\begin{figure}\vspace{-0.5cm}
\centering{\includegraphics[width=\linewidth]{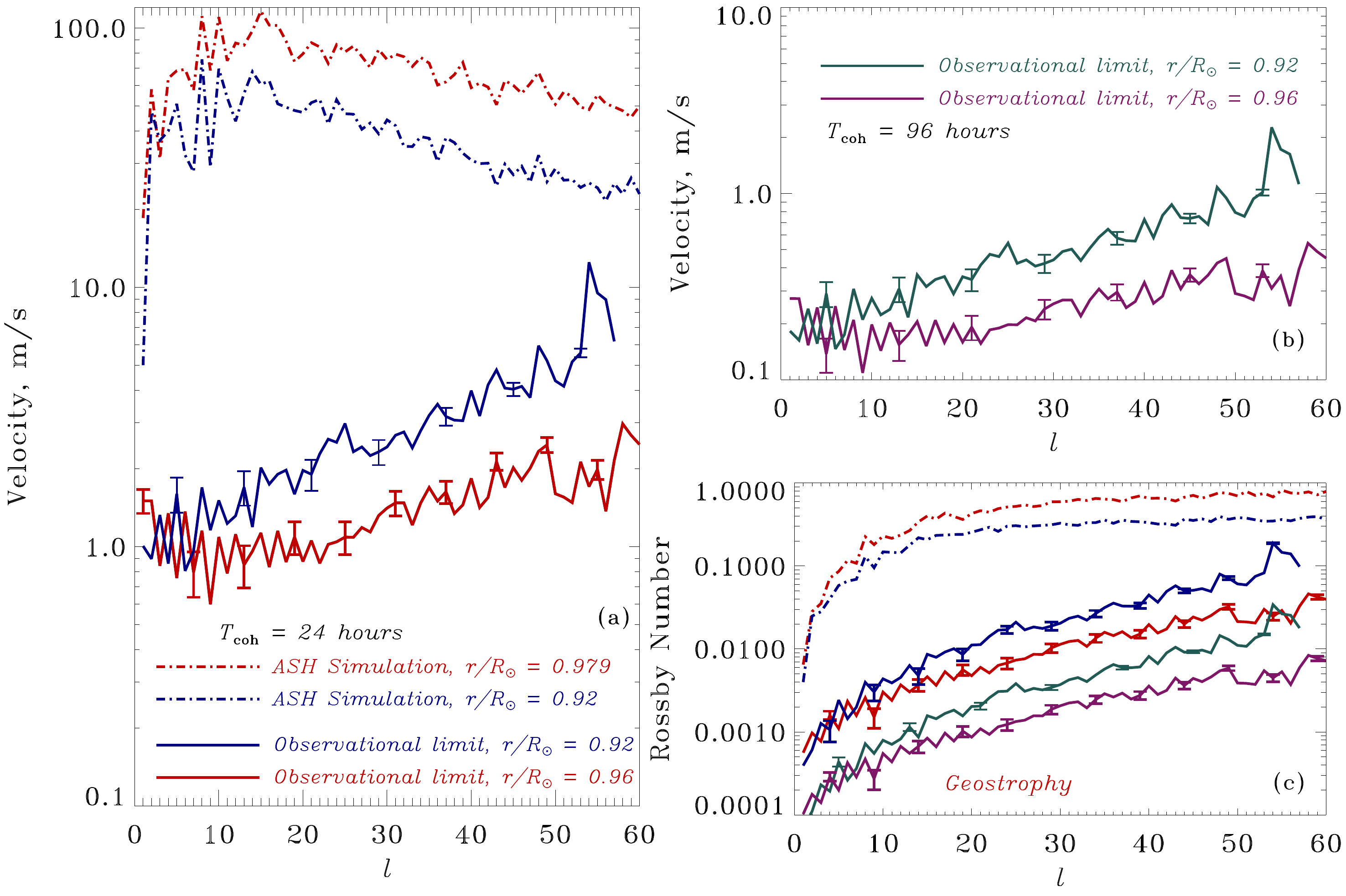}}\label{bounds}
\caption{
Observational bounds on flow magnitudes and the associated Rossby numbers.
Panels a, b: solid curves with 1-$\sigma$ error bars (standard deviations) show observational constraints on lateral flows averaged over $m$ at radial depths, $r/R_\odot = 0.92, 0.96$; dot-dash lines are spectra from ASH convection simulations \cite{miesch_etal_08}. Colours differentiate between the focus depth of the measurement and coherence times. At a depth of $r/R_\odot = 0.96$,
simulations of convection \cite{miesch_etal_08} show a coherence time of $T_{\rm coh} = 24$ hours (panel a) while
MLT \cite{spruit74} gives $T_{\rm coh} = 96$ hours (panel b), the latter obtained by dividing the mixing length by the predicted velocity. Both MLT and simulations \cite{nordlund09, trampedach11}
indicate a convective depth coherence over 1.8 pressure scale heights, an input to our inversion. At $r/R_\odot = 0.96$, MLT predicts a  60 ${\rm m\, s}^{-1}$, $\ell =61$ convective flow and for
$r/R_\odot = 0.92$, an $\ell = 33$, $45 \,{\rm m\, s}^{-1}$ flow (upon applying continuity considerations \cite{nordlund09}).
Panel c shows upper bounds on Rossby number, $Ro = U/(2\Omega L),\, L = 2\pi r/\sqrt{\ell(\ell+1)},\, r=0.92, 0.96 R_\odot$. Interior convection appears to be strongly geostrophically balanced (i.e., rotationally dominated) on these scales. By construction, these measurements are sensitive to lateral flows i.e., longitudinal and latitudinal at these specific depths ($r/R_\odot = 0.92, 0.96$) and consequently, we denote these flow components (longitudinal or latitudinal) by scalars.}

\end{figure}

Spatial scales on spherical surfaces are well characterized in spherical harmonic space: 
\begin{equation}
{\delta\tau}_{\ell m}(T) = \int_0^\pi \sin\theta\, d\theta \int_0^{2\pi} d\phi\,\, {\delta\tau}(\theta,\phi,T)\,Y^*_{\ell m}(\theta,\phi), 
\end{equation}
where $Y_{\ell m}$ are spherical harmonics, $(\ell,m)$ are spherical harmonic degree and order, respectively, and ${\delta\tau}_{\ell m}(T)$ are spherical harmonic coefficients.
Here, we specifically define the term ``scale" to denote $2\pi R_\odot/\sqrt{\ell(\ell +1)}$, which implies that small scales correspond to large $\ell$ and vice versa.
Note that a spatial ensemble of small convective structures such as a granules or inter-granular lanes (e.g., as observed on the solar photosphere) can lead to a broad power spectrum that has both small scales and large scales.
The power spectrum of an ensemble of small structures, such as granulation patterns seen at the photosphere, leads to
a broad distribution in $\ell$, which we term here as {\it scales}. %Small scales occur at high $\ell$ and vice versa. A small {\it structure} has a broad spectrum while a large feature
%is spectrally localized.  the power at large scales (i.e., low $\ell$) comprises contributions from small and large structures alike.
Travel-time shifts $\delta{\mathcal S}_{\ell m}$, induced by a convective flow component ${v}_{\ell m}(r)$, are given in the single-scattering limit by $\delta{\mathcal S}_{\ell m} = \int_\odot r^2\,dr \,{\mathcal K}_{\ell}(r)\,{v}_{\ell m}(r)$, where ${\mathcal K}_{\ell}$ is the sensitivity of the measurement to that flow component. The variance of flow-induced time shifts at every scale is bounded by the variance of the signal in observed travel times, i.e., $\langle\delta{\mathcal S}_{\ell m}^2\rangle \le {S}^2/\sigma^2(T_{\rm coh})\,\langle\delta\tau_{\ell m}^2(T_{\rm coh})\rangle$. To complete the analysis, we derive sensitivity kernels ${\mathcal K}_\ell(r)$ that allow us to deduce flow components in the interior, given the associated travel-time shifts (i.e., the {\it inverse} problem). %This sensitivity kernel is obtained by applying special analyses \cite{duvall06,hanasoge07} to numerical simulations \cite{hanasoge1,hanasoge_thesis} of waves propagating through flows in a solar-like 3D spherical shell (see Figure~\ref{kernels}).

The time-distance deep-focusing measurement \cite{hanasoge10} is calibrated by linearly simulating waves propagating through spatially small flow perturbations, implanted at 500 randomly distributed (known) locations, on a spherical shell at a given interior depth (Figure~\ref{kernels}). This delta-populated flow system contains a full spectrum, i.e., its power extends from small to large spherical harmonic degrees.
The simulated data are then filtered both spatially and temporally in order to isolate waves that propagate to the specific depth of interest (termed phase-speed filtering).
Travel times of these waves are then measured for focus depths the same as the depths of the features, and subsequently corrected for stochastic excitation
noise \cite{hanasoge07}. Note that these corrections may only be applied to simulated data - this is because we have full knowledge of the realization of sources that we put in.
Longitudinal and radial flow perturbations are analyzed through separate simulations, giving us access to the full vector sensitivity of this measurement to flows.
Travel-time maps from the simulations appear as a low-resolution
version of the input perturbation map because of diffraction associated with finite wavelengths of acoustic waves excited in the Sun and in the simulations. The connection between the two maps is primarily a function of spherical-harmonic degree $\ell$. To quantify the connection, both images are transformed and a linear regression is performed between coefficients of the two transforms
at each $\ell$ separately (see online supplementary material for details). The slope of this linear regression is the calibration factor for degree $\ell$.

We apply similar analyses to 27 days of data (one solar rotation) taken by the Helioseismic and Magnetic Imager
from June-July 2010. These images are tracked at the Carrington rotation rate, interpolated onto a fine latitude-longitude
grid, smoothed with a Gaussian, and resampled at the same resolution as the simulations (0.46875 deg/pixel). The data are transformed to spherical harmonic space and temporal Fourier domain, phase-speed filtered
(as described earlier) and transformed back to the real domain.
Cross correlations and travel times are computed with the same programs as used on the simulations.  Strips of 13 deg of longitude and the full latitude range are extracted from each of the 27 days'
results and combined into a synoptic map covering a solar rotation. The coefficients from the spherical harmonic transform of this map are converted, at each degree $\ell$, by the calibration slope mentioned above, and a resultant flow spectrum is derived, as shown in Figure~\ref{bounds}. These form observational upper bounds on the magnitude of turbulent flows in the convection zone
at the scales to which the measurements are sensitive.

It is seen that constraints in Figure~\ref{bounds} become poorer with greater imaging depth.
This may be attributed to diffraction, which limits seismic spatial resolution to approximately a wavelength. In turn, the acoustic wavelength, proportional to sound speed, increases with depth.
Since density also grows rapidly with depth, the velocity required to transport the heat flux of a solar luminosity decreases, a prediction echoed by all theories of solar convection.
Thus we may reasonably conclude that the $r/R_\odot = 0.96$ curve is also the upper bound for convective velocities at deeper layers in the convective zone (although the constraint
at $r/R_\odot = 0.92$ curve is weaker due to a coarser diffraction limit). Less restrictive
constraints obtained at depths $r/R_\odot = 0.79, 0.86$ (whose quality is made worse by the poor signal-to-noise ratio) are not displayed here.

%We have demonstrated that convective flow magnitudes in the Sun are 20-100 times weaker than simulations and mixing-length-theory
%predict, suggesting the prevalence of an intrinsically different paradigm of turbulent convection from that currently thought.
%We might thus ask: {\it How is the heat flux of a solar luminosity transported radially outward by motions that are so weak?} Figure~\ref{bounds} also shows that Rossby numbers are less than $10^{-2}$, almost 100 times smaller than simulations predict, suggesting that interior convection persists in a state of strong geostrophic balance. %Finally, the non-detection of
%convective structures at scales comparable to differential rotation casts doubt on the existence of a classical turbulent cascade.
%, in support of theories \cite{balbus09} that invoke {\it thermal wind balance} to explain solar differential rotation
\section{Discussion}
\subsection{Convective transport}
%The current understanding of turbulent transport of heat is that it is accomplished primarily by the small scales (e.g., \cite{miesch_etal_08}).
The spectral distribution of power due to an ensemble of convective structures, of spatial sizes small or large or both, will be broad. For example, it has
been argued (\cite{hathaway00}) that photospheric convection comprises only granules and supergranules, and that the power spectrum of
an ensemble of these structures would extend from the lowest to highest $\ell$.
In other words, if granulation-related flow velocities were to be altered, the {\it entire} power spectrum would be affected.
Thus the large scales which we image here (i.e., power for low $\ell$), contain contributions from small and large structures alike, and represent,
albeit in a complicated and incomplete manner, gross features of the transport mechanism.

Our constraints show that for wavenumbers $\ell < 60$, flow velocities associated with solar convection ($r/R_\odot = 0.96$) are substantially smaller than current predictions.
Alternately one may interpret the constraints as a statement that the temporal coherence of convective structures is substantially shorter than predicted by current theories. 
%Deeper within, i.e., $r/R_\odot < 0.96$, flows are likely to be even weaker due to the increasing density of the plasma.
%We claim this to be a challenge to the picture of convective transport and turbulence as put forth by e.g., \cite{miesch_etal_08}, who
%determine also
Analysis of numerical simulations (\cite{miesch_etal_08}) of solar convection shows that a dominant fraction ($\sim 80$\%) of the heat transport is effected by the small scales,
However, our observations show that the simulated velocities are substantially over-estimated in the wavenumber band $\ell < 60$, placing in question (based on the preceding argument)
the entire predicted spectrum of convective flows and the conclusions derived thereof. %The conclusions that one may derive from numerical simulations are based on the paradigm of turbulence
We further state that we lack definitive knowledge on the energy-carrying scales in the convection zone.
We may thus ask: how would this paradigm of turbulence affect extant theories of dynamo action?

For example, consider the scenario discussed by \cite{Spruit97}, who envisaged very weak upflows, which, seeded at the base of the convection
zone, grow to ever larger scales due to the decreasing density as they buoyantly rise. These flows are in mass balance with cool inter-granular plumes which, formed at the photosphere, are squeezed
ever more so as they plunge into the interior. Such a mechanism presupposes that these descending plumes fall nearly ballistically through the convection zone, almost as if a
cold sleet, amid warm upwardly diffusing plasma. In this schema, individual structures associated with the transport process would elude detection because the upflows would be too weak and the downflows of
too small a structural size (M. Sch\"{u}ssler, private communication, 2011). When viewed in terms of spherical harmonics, the associated velocities at large scales (i.e., low $\ell$), which contain contributions
from both upflows and descending plumes, would also be small. Whatever mechanism may prevail, the stability of descending plumes at high Rayleigh and Reynolds numbers and very low Prandtl number is
likely to play a central role (\cite{Spruit97, rast98}).

\subsection{Differential Rotation and Meridional Circulation}
Differential rotation, a large-scale feature ($\ell \sim 2$), is one individual global flow system and
easily detected in our travel-time maps. Differential rotation is the only feature we ``detect" within this wavenumber band. 
In other words, upon subtracting this $\ell =2$ feature from the travel-time maps, the variance of the remnant falls roughly as
$T^{-1}$, where $T$ is the temporal averaging length, suggesting the non-existence of other structures at these scales.  
Consequently, we may assert that we do not see evidence for a ``classical" inverse cascade that results in the production of
a smooth distribution of scales.

% The stability and amplitude of this feature induces travel-time shifts whose variance
%does not change with the amount of temporal averaging $T$. The same is not true at related wavenumbers (i.e., $\ell < 60$), where the average variance of time shifts
%falls roughly like $T^{-1}$. 

Current models of solar dynamo action posit that differential rotation drives the process of converting poloidal to toroidal flux. This
would result in a continuous loss of energy from the differentially rotating convective envelope and Reynolds' stresses have long been
thought of as a means to replenish and sustain the angular velocity gradient.
The low Rossby numbers in our observations indicate that turbulence is geostrophically arranged over
wavenumbers $\ell < 60$ at the depth $r/R_\odot = 0.96$, further implying very weak Reynolds stresses. Because flow velocities are likely to become weaker with depth in the convection zone, the
Rossby numbers will decrease correspondingly.
At wavenumbers of $\ell \sim 2$, the thermal wind balance equation describing
geostrophic turbulence likely holds extremely well within most of the convection zone:
\begin{equation}
\Omega_0 \frac{\partial\Omega}{\partial z} = \frac{C}{r^2\sin\theta}\frac{\partial S}{\partial\theta}, \label{geostrophy}
\end{equation}
where $\Omega_0$ is the mean solar rotation rate, $\Omega$ is the differential rotation, $z$ is the axis of rotation, $\theta$ is the latitude,
$C$ is a constant, $S$ is the azimuthally and temporally averaged entropy gradient.
Differential rotation around $\ell \sim 2$ is helioseismically well constrained, i.e., the left side of equation~(\ref{geostrophy}) is accurately known (e.g., \cite{kosovichev97}).
The iso-rotation contours are not co-aligned with the axis of rotation, yielding a non-zero left side of equation~(\ref{geostrophy}).
Taylor-Proudman balance is broken and we may reasonably infer that the Sun does indeed possess a latitudinal entropy gradient, of a suitable form so as to
sustain solar differential rotation (see e.g., \cite{kitchatinov95, balbus09}). 

The inferred weakness of Reynolds stresses poses a problem to theories of meridional circulation, which rely on the former to effect angular momentum transport
in order to sustain the latter. Very weak turbulent stresses would imply a correspondingly weak meridional circulation (e.g., \cite{rempel2005}).%, much smaller than the observed values of 10-20 ${\rm m\,s}^{-1}$.

%A great deal about the convection zone is unknown and the only means of discriminating various current theories is helioseismology. Inferences about details of the
%3-D interior structure and dynamics of the Sun prove both unexpected and revelatory.

\begin{acknowledgments}
All computing was performed on NASA Ames supercomputers: Schirra and Pleiades. SMH acknowledges support from NASA grant NNX11AB63G and thanks Courant Institute, NYU for hosting him as a visitor.
Many thanks to Tim Sandstrom of the NASA-Ames visualization group for having prepared Figure 1.
Thanks to M. Sch\"{u}ssler and M. Rempel for useful conversations. T. Duvall thanks the Stanford solar group for their hospitality. % and J. Schou for helpful comments.
Observational data that are used in our analyses here are taken by the Helioseismic and Magnetic Imager and are publicly available at http://hmi.stanford.edu/.
J. Leibacher and P. S. Cally are thanked for their careful reading of the manuscript and the considered comments that helped in improving it.
We thank M. Miesch for sending us the simulation spectra. The authors declare that they have no competing financial interests. Correspondence and requests for materials should be addressed to S.M.H.~(email: hanasoge@princeton.edu).
\end{acknowledgments}

%\bibliographystyle{pnas}
%\bibliography{convection}
%\end{article}
%\end{document}

\end{article}

% FIGURE 1

% FIGURE 2

% FIGURE 3

% FIGURE 4

% FIGURE 5
%
%Sincerely,

%Shravan Hanasoge, Thomas Duvall, Katepalli Sreenivasan

\end{document}